\documentclass{article}

\usepackage[utf8]{inputenc}

\usepackage{bbold}
\usepackage{microtype}
\usepackage{graphicx}
\usepackage{booktabs} 
\usepackage{amsthm}
\usepackage{amsfonts}
\usepackage{amsmath}
\usepackage{amssymb}
\usepackage{scrextend}
\usepackage{algorithm}
\usepackage{algorithmic}
\usepackage{subcaption}
\usepackage[english]{babel}
\usepackage{pgfplots}
\usepackage{multirow}
\usepackage{hyperref}
\usepackage{mathtools}
\usepackage{url}
\usepackage{comment}
\usepackage[numbers]{natbib}

\theoremstyle{plain}
\newtheorem{theorem}{Theorem}
\newtheorem{lemma}{Lemma}[section]

\newtheorem{corollary}{Corollary}[section]
\newtheorem{definition}{Definition}[section]
\pgfplotsset{compat=1.14}
\newcommand{\norm}[1]{\left\lVert#1\right\rVert}

\usepackage{amsmath}
\usepackage{amsfonts}
\usepackage{nicefrac}

\usepackage[accepted]{icml2021_style/icml2021}

\usepackage{todonotes}
\usepackage{fixme}
\fxsetup{status=final}
\FXRegisterAuthor{tv}{atv}{TV}

\begin{document}

\twocolumn[
\icmltitle{TEM: High Utility Metric Differential Privacy on Text}



\icmlsetsymbol{equal}{*}

\begin{icmlauthorlist}
	\icmlauthor{Ricardo Silva Carvalho}{sfu}
	\icmlauthor{Theodore Vasiloudis}{amz}
	\icmlauthor{Oluwaseyi Feyisetan}{amz}
\end{icmlauthorlist}

\icmlaffiliation{sfu}{Simon Fraser University}
\icmlaffiliation{amz}{Amazon.com}

\icmlcorrespondingauthor{Ricardo Silva Carvalho}{rsilvaca@sfu.ca}
\icmlcorrespondingauthor{Theodore Vasiloudis}{thvasilo@amazon.com}

\icmlkeywords{Differential Privacy, NLP}

\vskip 0.3in
]

\printAffiliationsAndNotice{}


\begin{abstract}
Ensuring the privacy of users whose data are used to train Natural Language Processing (NLP)
models is necessary to build and maintain customer trust. Differential Privacy (DP) has emerged
as the most successful method to protect the privacy of individuals.
However, applying DP to the NLP domain comes with unique challenges. The most successful
previous methods use a generalization of DP for metric spaces, and apply the privatization
by adding noise to inputs in the metric space of word embeddings. However,
these methods assume that one specific distance measure is being used,
ignore the density of the space around the input,
and assume the embeddings used have been trained on non-sensitive data.
In this work we propose Truncated Exponential Mechanism (TEM), a general method that allows the privatization
of words using any distance metric, on embeddings that can be trained on sensitive data.
Our method makes use of the exponential mechanism to turn the privatization step
into a \emph{selection problem}. 
This allows the noise applied to be calibrated to the
density of the embedding space around the input, and makes domain adaptation possible for the embeddings.
In our experiments, we demonstrate that our method significantly outperforms the state-of-the-art in terms
of utility for the same level of privacy, while providing more flexibility in the metric selection.


\end{abstract}

\section{Introduction}

Nowadays, text data are being used as input for a wide variety of machine learning tasks, from next word prediction in mobile keyboards \cite{hard2018federated}, to critical tasks like predicting patient health conditions from clinical records \cite{yao2019clinical}. 
Researchers have demonstrated that simple exploratory analysis tasks \cite{violate2004data} or the use of models trained on sensitive data \cite{shokri2017membership} may breach the privacy of the individuals involved. 
Even though there has been research focused on specific privacy-preserving tasks with textual data, such as language models \cite{mcmahan2017learning, bo2019er, vu2019dpugc}, to the best of our knowledge only a more recent line of work \cite{natasha2018obfuscation, poincareMDP2019, madlib2020} has been focusing on providing quantifiable privacy guarantees over the text itself.

In this work we aim to provide privacy guarantees to textual data using the formal notion of differential privacy (DP) \cite{dwork2006calibrating}, one of the most widely adopted privacy frameworks in industry and academia. Specifically, we apply a generalization called \textit{metric}-differential privacy \cite{chatzikokolakis2013metricdp}, which allows analysts to tailor solutions over general distance metrics. Previous work \cite{natasha2018obfuscation, poincareMDP2019, madlib2020} in this setting considered each input word by its vector representation and added noise to provide privacy guarantees. Additionally, as a noisy vector is unlikely to exactly represent a valid word, these methods returned a nearest neighbor approximation after querying the representation space. However, 
these works consider the representation space as non-sensitive, as they do not account for privacy loss in the nearest neighbor search. Moreover, the noise added to a vector does not take into account the density of the region that the vector lies in, which can potentially reduce the utility of a DP algorithm.


Our main contribution is the design of a new mechanism which we call the \textit{Truncated Exponential Mechanism} (TEM,) that satisfies metric-DP over textual data, posing the task as a selection problem. Instead of perturbing a representation vector, our method selects an output from a set of possible candidates where words closer to the input word in the metric space have higher probability of being selected. TEM works by adapting the probabilities of words being selected to the regions of a given input word, adjusting the noise injected for better utility, and allows the application of any formal distance function as metric. The mechanism includes a formal construction with a truncation step to initially select from high utility words with high probability, providing computationally efficient word selection with a tunable error parameter. Our experiments show that TEM obtains higher utility when compared to the state-of-the-art, for the same level of privacy.








\section{Related Work}\label{sec:rel_work}

\citet{natasha2018obfuscation} worked on text data via the ``bag of words'' representation of documents, and applied the Earth Mover's metric to obtain privatized bags, performing individual word privatization in the context of metric differential privacy. Following this context, the Madlib\footnote{We refer to this algorithm with the name used in \cite{madlibBlog}} mechanism \cite{madlib2020} adds noise to embedding vectors of words, working on in the Euclidean space and adding Laplacian noise to the embedding vectors. After introducing noise, the mechanism outputs the word that is closest to the noisy vector in the embedding space. The algorithm presented in \cite{poincareMDP2019} is a follow-up to \cite{madlib2020} although it appeared later. This mechanism works in a hierarchical embedding space, where the embedding vector of an input word is perturbed with noise from a hyperbolic distribution. 

\citet{poincareMDP2019} compare the hyperbolic mechanism to Madlib \cite{madlib2020}. However, since the two algorithms use different metric functions, the evaluation of privacy via only matching the $\varepsilon$ parameter of differential privacy can be improved. In this sense, \citet{poincareMDP2019} compares the privacy of the two mechanisms, looking at the probability of not changing a word after noise injection, i.e. the probability that the mechanism returns the exact same word used as input. Even though this notion can be intuitively seen as a level of indistinguishability, it cannot guarantee a fair comparison between mechanisms. The issue of comparing metric-DP mechanisms with different metric functions thus remains an open problem. In this work, we only compare mechanisms using the same metric function (Euclidean distance) to ensure a fair comparison.

\section{Preliminaries}

Consider a user giving as input a word $w$ from a discrete fixed domain $\mathcal{W}$. For any pair of inputs $w$ and $w'$, we assume a distance function $d: \mathcal{W} \times \mathcal{W} \rightarrow \mathbb{R}_{+}$, in a given space of representation of these words. More specifically, we consider a word embedding model $\phi: \mathcal{W} \rightarrow \mathbb{R}^n$ will be used to represent words, and the distance function can be a valid metric applicable to the embedding vectors.

Our goal is to select a word from $\mathcal{W}$, based on a given input, such that the privacy of a user, with respect to this word choice, is preserved. From an attacker's perspective, the output of an algorithm working over input $w$ or $w'$ will become more similar as these inputs become closer with respect to $d(w, w')$. 
Intuitively, words that are distant in metric space will be more easily distinguishable, compared to words that are close.

With that in mind, we will work on Metric-Differential Privacy \cite{chatzikokolakis2013metricdp}, a privacy standard defined for randomized algorithms with input from a domain $\mathcal{W}$ that are equipped with a distance metric $d: \mathcal{W} \times \mathcal{W} \rightarrow \mathbb{R}_{+}$ satisfying the formal axioms of a metric. In this context, the privacy guarantees given by metric-DP depend not only on the privacy parameter $\varepsilon$, but also the distance metric $d$ used. 

\begin{definition} (Metric Differential privacy \cite{chatzikokolakis2013metricdp})\label{def:metric_dp}. Given a distance metric $d: \mathcal{W} \times \mathcal{W} \rightarrow \mathbb{R}_{+}$, a randomized mechanism $\mathcal{M}: \mathcal{W} \rightarrow \mathcal{Y}$ is $\varepsilon d$-differentially private if for any $w, w' \in \mathcal{W}$ and all outputs $y \in \mathcal{Y}$ we have:
	\begin{gather}\label{eq:mDP}
	\Pr[\mathcal{M}(w) = y] \leq e^{\varepsilon d(w, w')} \Pr[\mathcal{M}(w') = y]
	\end{gather}
\end{definition}


For the Euclidean distance metric, as discussed on Section~\ref{sec:rel_work}, the current state-of-the-art is the Madlib mechanism, which adds Laplacian noise to a given vector in order to obtain a private output.

\begin{algorithm}[h]
	\caption{- \textbf{Madlib}: Word Privatization Mechanism for Metric Differential Privacy}\label{alg:madlib}
	\textbf{Input:} Finite domain $\mathcal{W}$, input word $w \in \mathcal{W}$ and privacy parameter $\varepsilon$.\;\\
	\textbf{Output:} Privatized element.
	\begin{algorithmic}[1]
		\STATE Compute embedding $\phi_w = \phi(w)$
		\STATE Perturb embedding to obtain $\hat{\phi}_w = \phi_w + N$ with noise density $p_N(z) \propto \exp(-\varepsilon \norm{z})$
		\STATE Return perturbed word $\hat{w} = argmin_{y \in \mathcal{W}} ||\phi(y) - \hat{\phi}_w||$
	\end{algorithmic}
\end{algorithm}

For a Euclidean metric $d: \mathcal{W} \times \mathcal{W} \rightarrow \mathbb{R}_{+}$, Madlib provides metric differential privacy.
	
\begin{theorem}\label{th:madlib} For a Euclidean metric $d$, Algorithm~\ref{alg:madlib} is $\varepsilon d$-differentially private.
\end{theorem}

Next we describe our algorithm that satisfies metric-DP, giving formal proof of its privacy guarantees.

\section{Metric Truncated Exponential Mechanism}

At its core, our algorithm uses the Exponential Mechanism (EM) \cite{mt2007expmech}, which is often used for selection in the context of differential privacy\cite{dwork2006calibrating}.

\begin{algorithm}[h]
	\caption{- \textbf{TEM}: Metric Truncated Exponential Mechanism}\label{alg:metric_EM}
	\textbf{Input:} Finite domain $\mathcal{W}$, input word $w \in \mathcal{W}$, truncation threshold $\gamma$, metric $d_{\mathcal{W}}: \mathcal{W} \times \mathcal{W} \rightarrow \mathbb{R}_{+}$, and privacy parameter $\varepsilon$.\;\\
	\textbf{Output:} Privatized element.
	\begin{algorithmic}[1]
		\STATE Given input $w$, obtain the set $L_{w}$ such that each word $w_i \in L_{w}$ satisfies $d_{\mathcal{W}}(w, w_i) \leq \gamma$
		\STATE Set the score $f(w, w_i)$ of each $w_i \in L_{w}$ as $f(w, w_i) = -d_{\mathcal{W}}(w, w_i)$
		\STATE Create a $\bot$ element with score $f(w, \bot) = -\gamma + 2\ln(| \mathcal{W} \setminus L_w  |)/\varepsilon$
		\STATE For each word $w_i \in L(x) \cup \bot$, add Gumbel noise with mean 0 and scale $2/\varepsilon$ to score $f(w, w_i)$
		\STATE Select $\hat{w}$ as the element with maximum noisy score from $L(x) \cup \bot$
		\STATE \textbf{if} $\hat{w} = \perp$ \textbf{then} return random sample of $\mathcal{W} \setminus L_w$
		\STATE \textbf{else} return $\hat{w}$
	\end{algorithmic}
\end{algorithm}

Algorithm~\ref{alg:metric_EM}, denoted as TEM, is using a variant of the exponential mechanism with Gumbel noise \cite{durfee2019practical}, and more specifically adapted to metric-DP for any given distance metric. TEM starts by selecting from words closer to the input by a distance of less or equal than a threshold $\gamma$, also including a $\bot$ element to account for the words outside the $\gamma$ distance. 
Privacy proof is deferred to Appendix~\ref{app:proof_th}.

\begin{theorem}\label{th:tem} For any formal distance metric $d$, Algorithm~\ref{alg:metric_EM} is $\varepsilon d$-differentially private.
\end{theorem}

To also optimize for utility, next we give a theoretical proposition for defining $\gamma$ in order to select elements within $\gamma$ distance from the input with high probability. We defer proof to Appendix~\ref{app:proof_utility}.

\begin{theorem}\label{th:tem_utility} For $\beta > 0$ and input $w \in \mathcal{W}$, TEM outputs elements with distance less or equal than $\gamma$ from $w$ with probability at least $1-\beta$ for $\gamma \geq \frac{2}{\varepsilon} \cdot \ln \frac{(1-\beta) (| \mathcal{W} | - 1)}{\beta}$.
\end{theorem}

The result above gives a guarantee of outputting words that are close to the input $w$ with high probability for a given $\gamma$ distance threshold. Thus, it is a theoretical way to choose $\gamma$ without looking at the data, i.e. without incurring privacy loss, while getting utility guarantees with high probability.

Below we highlight some of the advantages of TEM, specifically compared to the state-of-the-art.

\subsection{Detailed comparison with previous work}

TEM works for any given distance function that satisfies the axioms of a metric. In this sense, it has advantages for future use, when compared to previous mechanisms that used fixed metrics, such as Euclidean \cite{madlib2020} and Hyperbolic \cite{poincareMDP2019}, where changing the metric would need additional privacy analysis.

Previous work \cite{natasha2018obfuscation, madlib2020, poincareMDP2019} considered the text privacy preservation problem mainly as a task of releasing a word embedding vector after perturbing with some noise. This means they add noise to each of the dimensions of the vector they aim to release, treating every word embedding vector the same way. In practice, this leads to adding the same amount of noise for any word in the embedding space, regardless of whether the word lies in a dense or sparse region.
In contrast, TEM preserves privacy by posing the task as a \textit{selection problem}, giving words closer to the input word a higher probability of being selected. Therefore, TEM has a more dynamic behaviour, adjusting the noise to the domain of selection. In practice, for a given $\varepsilon$, TEM will add less noise to regions with high density, and more noise to regions with low density, therefore offering better utility in high-density areas.

Moreover, previous work assumed the word embeddings used were trained on a separate dataset, distinct from the data being privatized. TEM does not have that requirement, providing the potential for further utility gains through the use of domain adaptation, i.e. fine-tuning public pre-trained embeddings on the target sensitive data \cite{plank2013embedding, jaech2016domain}.

In terms of computational cost, the bottleneck of previous work resides in the nearest neighbor search of the noisy embedding vector obtained from the input. TEM starts by getting the elements within distance $\gamma$ of the input, but instead of querying the nearest neighbor, it queries for neighbors within a given range. In this sense, both methods can rely on fast approximate nearest neighbors implementations that support both querying nearest and by range, such as \cite{faissJDH17}. Nonetheless, TEM is the only mechanism that, for a fixed domain, is able to pre-process and store the search results for a given $\gamma$. After this step, the range search cost becomes constant. For the interested reader, we include a rewriting of Algorithm~\ref{alg:metric_EM} with a pre-processing step on Appendix~\ref{app:eff_version}.

Next we empirically compare our mechanism with the state of the art Madlib \cite{madlib2020} mechanism.

\section{Experiments}

For fair comparison to Madlib we use TEM with Euclidean distance on a fixed embedding space from GloVe \cite{pennington2014glove}. Experiments use the IMDB reviews dataset \cite{imdbData}. More details deferred to Appendix~\ref{app:experiments}.

\textbf{Utility}: To evaluate the utility of the metric-DP mechanisms, we build sentiment classification models on training data privatized by each mechanism and the baseline trained on sensitive data, and compare the accuracy of the trained models on a test dataset.

\textbf{Privacy}: As both mechanisms use the Euclidean distance metric, their privacy guarantees are matched by using the same $\varepsilon$. Nonetheless, for illustration we include the results of a Membership Inference Attack (MIA) \cite{shokri2017membership}, which tries to infer the presence of observations used to train a given model based only on black-box access. Lower attack score is better, representing more privacy preservation.

\begin{figure}[h]
	\centering
	\begin{subfigure}{\columnwidth}
		\includegraphics[scale=0.5]{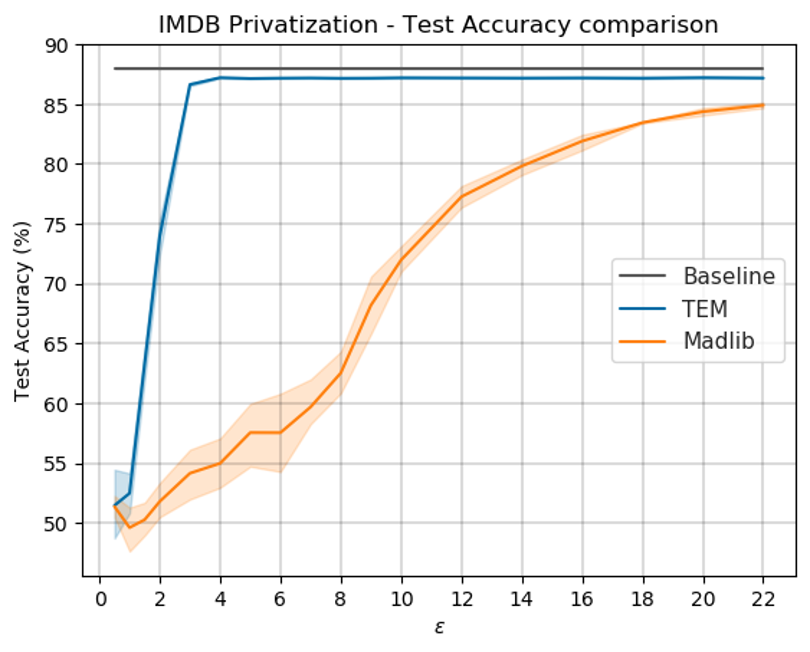}\centering
		\caption{Utility Evaluation}
		\label{fig:imdb_utility}
	\end{subfigure}%
	\\
	\begin{subfigure}{\columnwidth}
		\includegraphics[scale=0.5]{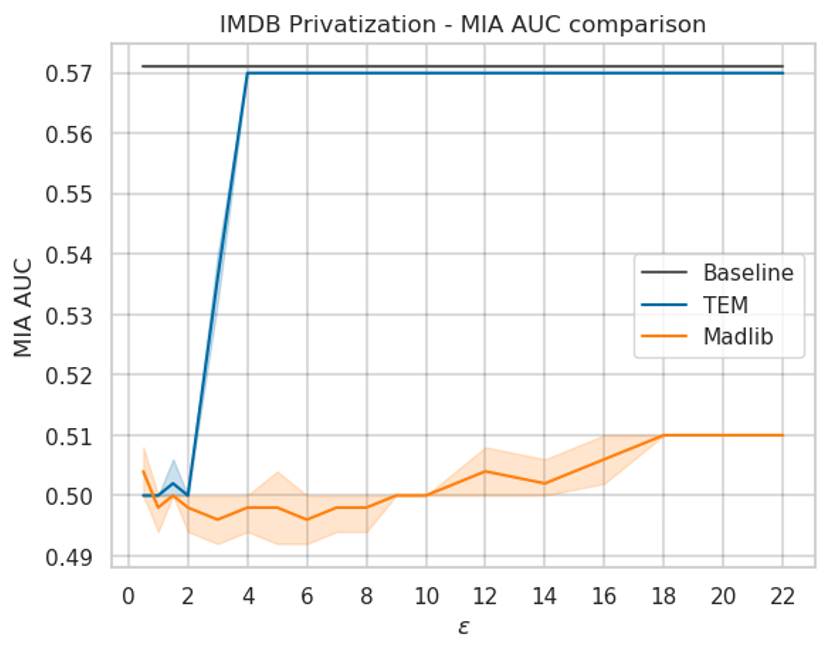}\centering
		\caption{Empirical Privacy Evaluation}
		\label{fig:imdb_mia}
	\end{subfigure}%
	\caption{Comparison of mechanisms with 95\% confidence interval over 5 trials for various $\varepsilon$. Baseline is built with models trained on original data. TEM used $\gamma$ from Theorem~\ref{th:tem_utility} with $\beta = 0.001$.}\label{fig:imdb}
\end{figure}

From the results of Figure~\ref{fig:imdb_utility} above we see that for fixed privacy level, TEM outperforms Madlib significantly. More specifically, we see from TEM's results that $\varepsilon \geq 4$ gives a formal level of privacy that is not significant. However, since Madlib adds more noise than needed for different regions of the embedding space, it still achieves some \textit{empirical} privacy, as observed by the MIA results on Figure~\ref{fig:imdb_mia}. Nonetheless, such enforcement is not a formal guarantee of privacy, which for metric-DP is bounded by $\varepsilon d(w, w')$, therefore in this context it is a loose guarantee of privacy for the embedding space considered. In this sense, when comparing TEM and Madlib for metric-DP formal levels of privacy, i.e. same $\varepsilon$ and metric space, we can clearly see better utility for TEM. Finally, as an example, if we look at $\varepsilon = 2$, where both mechanisms have $\text{AUC} = 0.50$ for MIA, we see Madlib with average test accuracy of $52\%$ and TEM with $75\%$, representing a relative utility improvement of $42\%$.


\section{Conclusion}

We presented TEM, a mechanism for text privatization on metric differential privacy with formal guarantees. Unlike the current state-of-the-art, our method allows the safe use of sensitive embeddings and provides flexibility of metric definition. In addition, TEM adapts the noise introduced to improve utility. Finally, it gives the possibility of performing pre-processing steps for enhanced computational efficiency. Our empirical evaluation demonstrates that TEM obtains better utility than the current state-of-the-art for the same formal privacy guarantees. As future work, we plan to perform domain adaptation, in order to leverage embeddings trained on sensitive data to improve utility. Including privatization of word context vectors is also a possible enhancement for improved accuracy.

\bibliographystyle{plainnat}
\bibliography{references.bib}

\clearpage

\appendix

\section{Omitted Proofs}\label{app:proofs}

In this section we include the omitted proofs.

\subsection{Proof of Theorem~\ref{th:tem}}\label{app:proof_th}

To prove the privacy of TEM, we need to first show that the sensitivity of the score function is still $d_{\mathcal{W}}(w,w')$ after the truncation. So first we prove a Lemma giving this result on the sensitivity, and after that we prove the privacy of our version of EM.

\begin{lemma}\label{lem:sensitivity} The sensitivity $\Delta f = \max\limits_{i \in \mathcal{W}} \max\limits_{w, w' \in \mathcal{W} } |f(i, w) - f(i, w')| \leq d_{\mathcal{W}}(w, w')$.
	
	\begin{proof} For a given input $w$, let us denote $\mathcal{I}(w)$ as the domain elements that have $d_{\mathcal{W}}(i, w) \leq \gamma$, therefore keeping their distances on the score function, while the elements on $\mathcal{W} \setminus \mathcal{I}(w)$ have distances fixed as $\gamma$.
		
		In this context, there are four possible cases we need to analyze for a given $i$ and any pair $w, w'$:, which we dive deep into now.
		
		\textbf{Case 1: $i \in \mathcal{I}(w)$ and $i \in \mathcal{I}(w')$}. If $i$ is in both $\mathcal{I}(w)$ and $\mathcal{I}(w')$, then it is using its original distance on the score for both $w$ and $w'$. Thus we have:
		\begin{gather}
		\nonumber f(i, w) - f(i, w') = -d_{\mathcal{W}}(i, w) + d_{\mathcal{W}}(i, w') \leq d_{\mathcal{W}}(w, w')
		\end{gather}
		with the last inequality being the use of the triangle inequality for the distance metric $d_{\mathcal{W}}$.
		
		\textbf{Case 2: $i \in \mathcal{I}(w)$ and $i \notin \mathcal{I}(w')$}. If $i$ is in $\mathcal{I}(w)$ but \textbf{not} in $\mathcal{I}(w')$, then it is using its original distance on the score for $w$ and $\gamma$ for $w'$, therefore:
		\begin{gather}
		\nonumber f(i, w) - f(i, w') = -d_{\mathcal{W}}(i, w) + \gamma
		\end{gather}
		Since $i$ is \textbf{not} in $\mathcal{I}(w')$, it means that $d_{\mathcal{W}}(i, w') > \gamma$, or equivalently $\gamma < d_{\mathcal{W}}(i, w')$, which replacing on the result above gives:
		\begin{gather}
		\nonumber f(i, w) - f(i, w') = -d_{\mathcal{W}}(i, w) + d_{\mathcal{W}}(i, w') \leq d_{\mathcal{W}}(w, w')
		\end{gather}
		where we use triangle inequality on the last step.
		
		\textbf{Case 3: $i \notin \mathcal{I}(w)$ and $i \in \mathcal{I}(w')$}. If $i$ is \textbf{not} in $\mathcal{I}(w)$ but is in $\mathcal{I}(w')$, then it is using $\gamma$ as the distance on score for $w$ and the original distance on score for $w'$, which gives us:
		\begin{gather}
		\nonumber f(i, w) - f(i, w') = -\gamma + d_{\mathcal{W}}(i, w')
		\end{gather}
		
		Since $i$ is in $\mathcal{I}(w')$, it means that $d_{\mathcal{W}}(i, w') \leq \gamma$, which replacing on the result above shows:
		\begin{align*}
			f(i, w) - f(i, w') &= -\gamma + d_{\mathcal{W}}(i, w') \\ 
			&\leq -\gamma + \gamma = 0 \leq d_{\mathcal{W}}(w, w')
		\end{align*}
		
		\textbf{Case 4: $i \notin \mathcal{I}(w)$ and $i \notin \mathcal{I}(w')$}. If $i$ is \textbf{not} in both $\mathcal{I}(w)$ and $\mathcal{I}(w')$, then it is using $\gamma$ as distance on score for both $w$ and $w'$, giving:
		\begin{gather}
		\nonumber f(i, w) - f(i, w') = -\gamma + \gamma = 0 \leq d_{\mathcal{W}}(w, w')
		\end{gather}
		
		Finally, we note that showing $f(i, w) - f(i, w') \geq - d_{\mathcal{W}}(w, w')$ on the same cases above follows by symmetry.
		
	\end{proof}
\end{lemma}

With the sensitivity result we can now show the privacy guarantee of the mechanism.

\textbf{Theorem~\ref{th:tem}.} \textit{For any formal distance metric $d$, Algorithm~\ref{alg:metric_EM} is $\varepsilon d$-differentially private.}

\begin{proof}
	
	For a given output $y \in \mathcal{W}$ and any pair of inputs $w, w' \in \mathcal{W}$ we have:
	\begin{gather}\label{eq:th_tem}
	\nonumber \frac{\Pr[M(w)=y]}{\Pr[M(w')=y]} =
	\\ \frac{\exp(\frac{\varepsilon}{2} \cdot f(y,w))}{\exp(\frac{\varepsilon}{2} \cdot f(y,w'))} \cdot \frac{ \sum\limits_{z' \in \mathcal{I}(w')} \exp(\frac{\varepsilon}{2} \cdot f(z',w')) + \sum\limits_{w' \in \mathcal{W} \setminus \mathcal{I}(w')} \exp(\frac{\varepsilon}{2} \cdot \gamma) }{ \sum\limits_{z \in \mathcal{I}(w)} \exp(\frac{\varepsilon}{2} \cdot f(z,w)) + \sum\limits_{w \in \mathcal{W} \setminus \mathcal{I}(w)} \exp(\frac{\varepsilon}{2} \cdot \gamma) }
	\end{gather}
	
	We note that on the second term above we have the same domain of elements, which even though they do not match among summations, they still satisfy the sensitivity on Lemma~\ref{lem:sensitivity}, which gives us for the first term on the right-hand side of Equation~\ref{eq:th_tem}:
	\begin{gather}
	\nonumber \frac{\exp(\frac{\varepsilon}{2} \cdot f(y,w))}{\exp(\frac{\varepsilon}{2} \cdot f(y,w'))} \leq 
	\\ \frac{\exp(\frac{\varepsilon}{2} \cdot ( f(y,w') + d_{\mathcal{W}}(w, w') ) )}{\exp(\frac{\varepsilon}{2} \cdot f(y,w'))} \leq \exp(\frac{\varepsilon}{2} \cdot d_{\mathcal{W}}(w, w') )
	\end{gather}
	
	And similarly for the second term on the right-hand side of Equation~\ref{eq:th_tem}: 
	\begin{gather}
	\nonumber \frac{ \sum\limits_{z' \in \mathcal{I}(w')} \exp(\frac{\varepsilon}{2} \cdot f(z',w')) + \sum\limits_{w' \in \mathcal{W} \setminus \mathcal{I}(w')} \exp(\frac{\varepsilon}{2} \cdot \gamma) }{ \sum\limits_{z \in \mathcal{I}(w)} \exp(\frac{\varepsilon}{2} \cdot f(z,w)) + \sum\limits_{w \in \mathcal{W} \setminus \mathcal{I}(w)} \exp(\frac{\varepsilon}{2} \cdot \gamma) } \leq \exp(\frac{\varepsilon}{2} d_{\mathcal{W}}(w, w'))
	\end{gather}
	
	Therefore, multiplying the two inequalities above gives us for Equation~\ref{eq:th_tem}:
	\begin{gather}
	\nonumber \frac{\Pr[M(w)=y]}{\Pr[M(w')=y]} \leq 
	\\ \exp(\frac{\varepsilon}{2} d_{\mathcal{W}}(w, w')) \cdot \exp(\frac{\varepsilon}{2} d_{\mathcal{W}}(w, w')) = \exp(\varepsilon d_{\mathcal{W}}(w, w'))
	\end{gather}
	which proves the metric-DP guarantee of the mechanism.
	
	The only difference in writing from TEM to what we used in the probabilities here is that instead of directly picking from all the elements with distance greater than $\gamma$ we use $\bot$ first. So now we show they are equivalent.
	
	Let $L(w)$ be a list of elements where each element $i$ satisfies $d_{\mathcal{W}}(i, w) \leq \gamma$ and let $\bar{L}(w)$ be the list of remaining elements $\mathcal{W} \setminus L(w)$. For elements in $\bar{L}(w)$, we see that on TEM they are selected randomly after $\bot$ is selected. Since $\bot$ has score $-\gamma + 2\ln(|\bar{L}(w)|)/\varepsilon$, on the exponential mechanism this is equivalent to:
	\begin{align*}
	 \exp(\varepsilon/2 \cdot (-\gamma + 2\ln(|\bar{L}(w)|)/\varepsilon) ) &=
	\\
	 \exp(\varepsilon/2 \cdot -\gamma) \cdot \exp(\varepsilon/2 \cdot ( 2\ln(|\bar{L}(w)|)/\varepsilon) ) &=
	\\
	 \exp(\varepsilon/2 \cdot -\gamma) \cdot \exp(\ln(|\bar{L}(w)|)) ) &=
	\\
	 \exp(\varepsilon/2 \cdot -\gamma) \cdot |\bar{L}(w)|
	\end{align*}
	
	This result is essentially the same as using $|\bar{L}(w)|$ elements with score $-\gamma$ directly for selection since, after selecting $\bot$, the elements in $\Bar{L}(w)$ are selected randomly, giving each a probability of selection proportional to $\exp(\varepsilon/2 \cdot -\gamma)$.
	
\end{proof}

\subsection{Proof of Theorem~\ref{th:tem_utility}}\label{app:proof_utility}

The following proof for Theorem~\ref{th:tem_utility} only uses the fact that on an exponential mechanism we have the probability of a given element proportional to the score of the element.

\textbf{Theorem ~\ref{th:tem_utility}.} \textit{For $\beta > 0$ and input $w \in \mathcal{W}$, TEM outputs elements with distance less or equal than $\gamma$ from $w$ with probability at least $1-\beta$ for $\gamma \geq \frac{2}{\varepsilon} \cdot \ln \frac{(1-\beta) (| \mathcal{W} | - 1)}{\beta}$.}

\begin{proof}
	This theorem is equivalent to guaranteeing that we only output elements outside the $\gamma$ distance of the input with probability at most $\beta$. The worst case to guarantee this condition is when we only have the input word inside the $\gamma$ distance, and all of the remaining $| \mathcal{W} | - 1 $ words are outside. Thus we calculate the probability in this theorem for this worst case to obtain the maximum guarantee.

\begin{align*}
	  \frac{ ( | \mathcal{W} | - 1 ) \cdot \exp(-\varepsilon/2 \cdot \gamma ) }{ \exp(-\varepsilon/2 \cdot 0 ) +  ( | \mathcal{W} | - 1 ) \cdot \exp(-\varepsilon/2 \cdot \gamma )} &\leq \beta
\\
	 \frac{ 1 }{ 1 / \big( ( | \mathcal{W} | - 1 ) \cdot \exp(-\varepsilon/2 \cdot \gamma ) \big) + 1} &\geq \beta
\\
	 1 / \big( ( | \mathcal{W} | - 1 ) \cdot \exp(-\varepsilon/2 \cdot \gamma ) \big) + 1 &\leq 1/\beta
\\
	 1 / \big( ( | \mathcal{W} | - 1 ) \cdot \exp(-\varepsilon/2 \cdot \gamma ) \big) &\geq (1-\beta)/\beta
\\
	 ( | \mathcal{W} | - 1 ) \cdot (1-\beta)/\beta  &\leq \exp(\varepsilon/2 \cdot \gamma )
\\
	 (2/\varepsilon) \cdot \ln ( ( | \mathcal{W} | - 1 ) \cdot (1-\beta)/\beta )  &\leq \gamma
\end{align*}

\end{proof}

\section{Efficient Version}\label{app:eff_version}

Here we precisely describe a version of our algorithm with exact search for the elements within $\gamma$ distance of any input.

A first glance at TEM reveals it can be computationally prohibitive for large domains, so below we develop a version of our mechanism that is computationally efficient leveraging the properties of our algorithm.
 
For a fixed finite domain $\mathcal{W}$, given a truncation threshold $\gamma$, we pre-compute, for each possible input, the list of elements that satisfy $d_{\mathcal{W}}(i, w) \leq \gamma$, which makes the search for possible candidates of a given input $O(1)$. Another simplification already included in TEM is to use a version of the Exponential Mechanism that uses Gumbel noise \cite{durfee2019practical}, which helps avoid dealing with probabilities using the exponential function. Finally, we also point out that we can group all elements below the truncation threshold as one element $\bot$ with the aggregated count, and then if $\bot$ gets selected, randomly sample one of the aggregated elements. Below we give such simplified algorithm and prove it is equivalent to Algorithm~\ref{alg:metric_EM}.

\begin{algorithm}[H]
	\caption{Metric Truncated Exponential Mechanism}\label{alg:metric_EM_fast}
	\textbf{Input:} Finite domain $\mathcal{W} = \{1, ..., m\}$ of elements, element index $x \in \mathcal{W}$, truncation threshold $\gamma$, metric $d_{\mathcal{W}}: \mathcal{W} \times \mathcal{W} \rightarrow \mathbb{R}$, and privacy parameter $\varepsilon$.\\
	\textbf{Output:} Element index.
	\begin{algorithmic}[1]
		\STATE \textbf{Pre-processing:}
		\FOR{each $x \in \mathcal{W}$}
		\STATE Create a list $L(w)$ of elements where each element $i$ satisfies $d_{\mathcal{W}}(i, w) \leq \gamma$ and let $\bar{L}(w)$ be the list of remaining elements $\mathcal{W} \setminus L(w)$.
		\STATE Define for each $x \in \mathcal{W}$ the score of a $\bot$ element as $f_{\bot}(w) = -\gamma + 2\ln(|\bar{L}(w)|)/\varepsilon$
		\ENDFOR
		\STATE \textbf{Selection:}
		\STATE Given input $x \in \mathcal{W}$, for every element in $L(w) \cup \bot$ add noise from a Gumbel distribution with mean 0 and scale $2/\varepsilon$ to each score $-d_{\mathcal{W}}(\cdot, x)$
		\STATE Set $y$ as the element with maximum noisy score
		\IF{$y = \perp$}
		\STATE Return random sample of $\bar{L}(w)$
		\ELSE
		\STATE Return $y$
		\ENDIF
	\end{algorithmic}
\end{algorithm}

We now formally prove Algorithm~\ref{alg:metric_EM_fast} is equivalent to Algorithm~\ref{alg:metric_EM}.

\begin{lemma} Algorithm~\ref{alg:metric_EM_fast} and Algorithm~\ref{alg:metric_EM} are equal in distribution.
	\begin{proof}
		
		The only difference between the algorithms is that on Algorithm~\ref{alg:metric_EM_fast} we pre-process $L(w)$ and $\bar{L}(w)$ ahead of time. Since for a fixed domain they do not change and are independent for each input word, the algorithms are equal in distribution.
		
	\end{proof}
\end{lemma}

Since the two algorithms are equivalent, Algorithm~\ref{alg:metric_EM_fast} also satisfies the same privacy guarantee.

\begin{corollary} Algorithm~\ref{alg:metric_EM_fast} is $\varepsilon d_{\mathcal{W}}$-DP.
\end{corollary}

\section{Experiments details}\label{app:experiments}

In this section, we describe the exact settings and include all of the details for the mechanisms used, to allow reproducibility.

We used the IMDB dataset  \cite{imdbData}, which gives two different files: training data and testing data, each with 25.000 examples. For the baseline we trained a model using dataset TR1 with 50\% of the IMDB training data and tested it on TE1 with 50\% of the IMDB testing data. For the privatized utility, we trained models on TR1 after privatization by each mechanism and tested them on original TE1.

For MIA the models trained as described above were attacked, we denote a given target model as T. For the shadow model, denoted as S, we trained a model on dataset TE2 having the other 50\% of the IMDB testing data. To train the attack model, denoted as A, we used as features the output of TE1 and TE2 given by S, where TE2 is labeled as "in" and TE1 as "out". After training model A, we evaluated the inference attack with TR1 and TR2 (having another 50\% of the IMDB training data) with features being the output of TR1 and TR2 obtained by a given target model T, where ground-truth for TR1 is "in" and for TR2 is "out.

For embeddings we used GloVe \cite{pennington2014glove} with 300 dimensions. The sentiment classification models follow the FastText classifier \cite{Joulin_2017}, whereas the attack model is an MLP with two-layers having 64 hidden nodes each, and ReLU activations. Each model was trained for 20 epochs with batch size of 64 and default pytorch parameters for Adam optimizer.

\end{document}